\documentstyle[12pt]{article}
\def\nabstar#1{\nabla\kern-0.5pt\smash{\raise 4.5pt\hbox{$\ast$}}
               \kern-4.5pt_{#1}}

\def\drvstar#1{\partial\kern-0.5pt\smash{\raise 4.5pt\hbox{$\ast$}}
               \kern-5.0pt_{#1}}

\def\newline{\relax\ifhmode\null\hfil\break\else\nonhmodeerr@\newline\fi}
\def\frac#1#2{{#1\over#2}}
\def\text#1{{\hbox{\rm #1}}}

\newcommand{\beq}{\begin{equation}}
\newcommand{\eeq}{\end{equation}}
\newcommand{\bea}{\begin{eqnarray}}
\newcommand{\eea}{\end{eqnarray}}
\def\Id{ \mbox{1\hspace{-1.2mm}I} }
\begin{document}
\thispagestyle{empty}
\begin{flushright}
NTUTH-98-060 \\
June 1998
\end{flushright}
\bigskip\bigskip\bigskip
\vskip 2.5truecm
\begin{center}
{\LARGE {Solutions of the Ginsparg-Wilson Relation}}
\end{center}
\vskip 1.0truecm
\centerline{Ting-Wai Chiu\footnote{E-mail address: twchiu@phys.ntu.edu.tw} 
and Sergei V. Zenkin
\footnote{Permanent address: Institute for Nuclear
Research of the Russian Academy of Sciences, 117312 Moscow, Russia.
E-mail address: zenkin@al20.inr.troitsk.ru}}
\vskip5mm
\centerline{Department of Physics, National Taiwan University} 
\centerline{Taipei, Taiwan 106, R.O.C}
\vskip 2cm
\bigskip \nopagebreak \begin{abstract}
\noindent

We analyze general solutions of the Ginsparg-Wilson relation 
for lattice Dirac operators and formulate a necessary condition
for such operators to have non-zero index in the topologically
nontrivial background gauge fields.

\bigskip
\bigskip
\bigskip


\end{abstract}
\vskip 1.5cm

\newpage\setcounter{page}1

Recently there are very interesting developments in theoretical understandings
of the chiral symmetry on the lattice. The idea stems from the
Ginsparg-Wilson (GW) relation \cite{gwr} which was derived in 1981 as the
remnant of chiral symmetry on the lattice after blocking a chirally
symmetric theory with a chirality breaking local renormalization group
transformation. The original GW relation is
\beq
D \gamma_5 + \gamma_5 D = 2 a D \gamma_5 R D,
\label{eq:gwo}
\eeq
where $ D $ is lattice Dirac operator, $ R $ is a non-singular hermitian
operator which is local in the position space and trivial in the Dirac space,
and $ a $ is the lattice spacing which reminds us the fact that $ D $
becomes chirally symmetric in the continuum limit $ a \rightarrow 0 $.
According to the Nielsen-Ninomiya theorem \cite{no-go} the chiral
symmetry of a local Dirac operator defined on the regular lattice
must be broken in order to avoid the species doubling.
The main advantage of the GW relation is that it introduces
the chiral symmetry breaking of $D$ in the mildest way \cite{gwr}.
Although it does not ensure the absence of the species
doubling, it does incorporate two remarkable properties in the following.

The first is that the action $ A = \bar\psi D \psi $ has an exact symmetry :
\bea
\label{eq:lus1}
\psi &\rightarrow& \exp [ i \theta  \gamma_5 ( \Id - R D ) ] \psi, \\
\label{eq:lus2}
\bar\psi &\rightarrow& \bar\psi \exp [ i \theta (\Id- D R) \gamma_5 ],
\eea
where $ \theta $ is a global parameter, which was
discovered by L\"uscher \cite{ml98:2}.
The second is that any operator $D$ satisfying the GW relation possesses a
well defined integer index on a finite lattice \cite{ph98:1}, \cite{ml98:2}
\beq
\lim_{\epsilon \rightarrow 0} \epsilon \sum_n \langle  \overline{\psi}_n 
\gamma_5 \psi_n \rangle_f = \mbox{Tr}(\gamma_5 R D) = n_- - n_+ \equiv
\mbox{index }D,
\label{ind}
\eeq
where l.h.s.\ stands for the fermionic average of $\overline{\psi}
\gamma_5 \psi$ calculated with the infinitesimal mass $\epsilon$ added to the
operator $D$, and $n_{+}$ ($n_-$) are the number of the zero modes of
$D$ with positive (negative) chiralities.
This is in contrast to the Wilson-Dirac operator for which the l.h.s.\
generally is not an integer on a finite lattice.

It is essentially due to these two properties that such formulations of
lattice QCD can possess the attractive features pointed out in
\cite{ml98:2}-\cite{ph98:2}.
However only the GW relation itself is not sufficient to
guarantee that any $ D $ satisfying (\ref{eq:gwo}) must possess exact
zero modes with definite chiralities, and reproduce the Atiyah-Singer
index theorem on the lattice. In this paper, we analyze general solutions
of GW relation and formulate a necessary condition for them to have
non-zero indices in topologically non-trivial background gauge fields. 
We limit our consideration to the operators $D$ satisfying the
hermiticity property
\beq
D^{\dagger} = \gamma_5 D \gamma_5 .
\label{eq:herm}
\eeq

First, we consider the case of non-singular $D$ which is relevant
to topologically trivial gauge field background, except possibly some 
`exceptional' configurations. Then eq.\ (\ref{eq:gwo}) is equivalent to
the following equation linear in $ D^{-1} $,
\beq
\gamma_5 D^{-1} + D^{-1} \gamma_5 = 2  a \gamma_5 R,
\label{eq:gwi}
\eeq
and its general solution can be written in the form 
\beq
\label{eq:gen_soln}
D = (\Id + a D_c R )^{-1} D_c = D_c ( \Id + a R D_c )^{-1}
\eeq
where $ D_c $ is the chirally symmetric lattice Dirac operator, i.e.
\beq
D_c \gamma_5 + \gamma_5 D_c = 0
\label{eq:ch}
\eeq
Thus in the nonsingular case the problem of constructing explicit solutions of
$D$ reduces to finding a proper realization of the chirally symmetric operator
$D_c$. Note that by virtue of the condition (\ref{eq:herm}) and
eq.\ (\ref{eq:ch}) the operator $D_c$ is antihermitian, and therefore, normal.
In order to avoid species doubling for $D$ defined on a regular lattice,
$D_c$ should be non-local. Additional limitations to the form of $D_c$
come from the requirement of the locality of $D$. For a more detailed
discussion on the properties of $D_c$ we refer to our paper \cite{CWZ}
where a few explicit examples are also presented.

It is interesting to observe that in this case both $ D_c $ and $ D $ can
be constructed from a
unitary operator $ V $ ( $V^{\dagger} = V^{-1}$ ) which satisfies
the hermiticity condition
\beq
\label{eq:herm_V}
\gamma_5 V \gamma_5 = V^{\dagger}
\eeq
Indeed, for any given chirally invariant $ D_c $ satisfying 
(\ref{eq:herm}), $ D_c $ is antihermitian, so that
$ \mbox{det}(a D_c + \Id ) \ne 0 $, then there exists a unitary operator
\beq
\label{eq:UV}
V = ( a D_c - \Id)( a D_c  + \Id )^{-1}
\eeq
satisfying (\ref{eq:herm_V}). So $ D_c $ can be
represented as
\beq
D_c = a^{-1} \frac{\Id+V}{\Id-V},
\label{eq:DcV}
\eeq
provided that $\Id - V$ is nonsingular.
Substituting (\ref{eq:DcV}) into (\ref{eq:gen_soln}),
we obtain the general solution of (\ref{eq:gwi}) for nonsingular $ D $ in 
terms of the unitary operator $V$
\bea
\label{eq:DV}
D &=& a^{-1} [ (\Id + V ) R + \Id-V ]^{-1} (\Id + V )  \\
  &=& a^{-1} (\Id+V)[ R (\Id+V) + \Id-V ]^{-1}
\label{eq:DV1}
\eea
Note that in contrast to eq.\ (\ref{eq:DcV}) these expressions make sense
even when operator $\Id - V$ is singular.
As we will show later, due to this fact eqs.\ (\ref{eq:DV}) and (\ref{eq:DV1})
represent a class of solutions of (\ref{eq:gwo}) also for singular $D$, and
thus are valid for any gauge configurations.

Let us now consider the case when $D$ is singular, i.e.\ $\mbox{det}D =0$,
which is the result we would like to have in the topologically non-trivial
gauge field background. In this case we are interested in
only those $D$ which have the possibility to reproduce the index theorem on 
the lattice, i.e.\ have a non-zero index in (\ref{ind}). So we obtain a
necessary condition for any solutions of (\ref{eq:gwo}) to possess
non-zero indices in topologically non-trivial background gauge fields.
We hope that this condition not only serves as a discriminant to rule
out any unphysical solutions of GW relation but also can provide guidelines
to construct viable solutions of GW relation.
We state our result in the following theorem.

{\em Theorem} : For any lattice Dirac operator
$ D $ satisfying hermiticity condition (\ref{eq:herm}) and the
GW relation (\ref{eq:gwo}), the necessary condition for it
to have non-zero index in the topologically nontrivial gauge background is
\beq
\label{eq:det1}
\mbox{det} (\Id - a D R ) = 0
\eeq
or, equivalently
\beq
\label{eq:det2}
\mbox{det} (\Id - a R D ) = 0
\eeq

{\em Proof } : Assume that $ \mbox{det}( \Id - a D R ) \ne 0 $.
Then there exists a chirally symmetric normal operator
\beq
\label{eq:Dc}
D_c = ( \Id - a D R )^{-1} D. 
\eeq
It is obvious that any zeromode of $D$ is a zeromode of $D_c$, and vice versa.
Therefore $\mbox{index } D_c = \mbox{index } D$. However according to the
theorem proved in \cite{zenkin98:3}, the index of any chirally symmetric
normal Dirac operator is zero,
so $\mbox{index } D_c = \mbox{index } D = 0$.
Since $ \mbox{det}( \Id - a D R )= \mbox{det}( \Id - a R D )$,
this completes the proof.

In other words, we have proved that in order the operator $D$ to have nonzero
index the chirally invariant operator $D_c$ in (\ref{eq:Dc}), and therefore in
(\ref{eq:gen_soln}), should not exist. However $ D $ is still well
defined and exists. Eqs.\ (\ref{eq:DcV}) and (\ref{eq:DV}) suggest a simple
interpretation of this seemingly paradoxial situation.
As shown in \cite{twc98:4}, for any unitary operator $ V $ satisfying
(\ref{eq:herm_V}), if it has real ( $ \pm 1 $ ) eigenmodes
then these real eigenmodes are chiral and
the total chirality of all real eigenmodes must vanish
\beq
\label{eq:totchi}
  n_{+1}^{+} - n_{+1}^{-} + n_{-1}^{+} - n_{-1}^{-} = 0
\eeq
where $ n_{+1}^{+} $ ( $ n_{+1}^{-} $ ) denotes the number of positive
( negative ) chirality eigenmodes of eigenvalue $ +1 $, while
$ n_{-1}^{+} $ ( $ n_{-1}^{-} $ ) denotes the number of positive
( negative ) chirality eigenmodes of eigenvalue $ -1 $.
The $ -1 $ eigenmodes of $ V $ correspond to the zero modes of $ D $.
Thus, if $D$ has non-zero index ( $ n_{-1}^{-} - n_{-1}^{+} \ne  0 $ ),
then $ n_{+1}^{+} - n_{+1}^{-} \ne  0 $ and $ V $ has eigenvalue $ +1 $.
Then the chirally invariant operator $D_c$ in (\ref{eq:DcV}) is no longer
defined, while the operator $D$ in (\ref{eq:DV}) becomes singular
but still well defined. Therefore (\ref{eq:DV})
is indeed a class of general solutions for the GW relation (\ref{eq:gwo})
for any gauge configurations.

Finding a unitary operator $V$
which can have eigenvalues $ +1 $ and $ -1 $ in the
topologically non-trivial sectors,
however, is a highly non-trivial task. So far we
know only one explicit example of $V$ which does satisfy this requirement.
It is the unitary operator
derived from the overlap formalism \cite{hn97:7},
\beq
\label{eq:V}
V = D_w ( D_w^{\dagger} D_w )^{-1/2},
\eeq
where $ D_w $ is the standard Wilson-Dirac
operator but with a negative mass in the range $ (- 2 a^{-1}, 0) $.
In \cite{twc98:4}, it has been demonstrated that this solution of GW relation
indeed reproduces exact zero modes and the
index theorem is satisfied exactly on a finite lattice. The zero modes are
also in very good agreement with the continuum theory.
At this moment we cannot provide another example of $ V $
which can satisfy all our requirements.

It is instructive to note that solutions of the GW relation may have zero
index not only because zero modes with opposite chiralities always appear in
pairs but also because they may not have any zero modes at all.
Consider the naive massless Dirac fermion operator $ D_n $ 
on the regular lattice, the random lattice \cite{cfl82}
and the random-block lattice \cite{twc88} respectively.
Since $ D_n $ is chirally symmetric, it can be taken as $ D_c $ and the
Dirac operator $ D $ can be constructed from (\ref{eq:gen_soln}).
For any one of these GW-Dirac operators, we do not find any genuine zero
modes in any topologically non-trivial sectors.

To summarize, we have demonstrated that the GW relation does not guarantee the
existence of exact zero modes nor the realization of index theorem on the
lattice. If a solution of GW relation, $ D $, in topologically nontrivial 
sector gives
$ \mbox{det}( \Id - a R D ) \ne 0 $, and therefore can be expressed in terms 
of a chirally invariant operator $D_c$,
its index must be zero, and thus it should be dropped from the
list of viable lattice Dirac fermion operators.
We note in passing that the necessary condition
(\ref{eq:det1}) or (\ref{eq:det2}) is precisely the condition that the
generators of the lattice chiral transformations in (\ref{eq:lus1}) and
(\ref{eq:lus2}) are singular.
The general solution for the GW relation (\ref{eq:gwo})
is obtained in (\ref{eq:DV}).

\bigskip
\bigskip

\vfill\eject

{\bf Acknowledgement }
\bigskip

This work was supported by the National Science Council, R.O.C. under
the grant number NSC87-2112-M002-013. \
T.W.C. would like to thank Herbert Neuberger for enlightening
correspondences. \
S.V.Z is grateful to members of the Department of Physics at National
Taiwan University for the hospitality extended to him during his stay
at Taipei.

\vfill\eject

\end{document}